\documentstyle[aps,prl,floats,epsf]{revtex}
\begin{document}
\advance\textheight by 0.2in
\draft

\twocolumn[\hsize\textwidth\columnwidth\hsize\csname@twocolumnfalse%
\endcsname

\title{Multicritical Behavior in Coupled Directed Percolation
        Processes} 

\author{Uwe C. T\"auber$^{1,*}$, Martin J. Howard$^2$, and 
        Haye Hinrichsen$^3$}

\address{$^1$ Department of Physics --- Theoretical Physics,
        University of Oxford, 1 Keble Road, Oxford OX1 3NP \\ 
        and Linacre College, St. Cross Road, 
        Oxford OX1 3JA, United Kingdom \\
        $^2$ CATS, The Niels Bohr Institute, Blegdamsvej 17,
        2100 Copenhagen \O, Denmark \\
        $^3$ Max-Planck-Institut f\"ur Physik komplexer Systeme,
        N\"othnitzer Stra\ss e 38, D-01187 Dresden, Germany \\ } 

\date{\today, submitted to Phys. Rev. Lett., OUTP-97-24-S}

\maketitle

\begin{abstract}
We study a hierarchy of directed percolation (DP) processes for
particle species $A, B, \ldots$, unidirectionally coupled via 
the reactions $A \to B, \ldots$. 
When the DP critical points at all levels coincide, multicritical
behavior emerges, with density exponents $\beta^{(k)}$ which are
markedly reduced at each hierarchy level $k \geq 2$.
We compute the fluctuation corrections to $\beta^{(2)}$ to 
$O(\epsilon = 4-d)$ using field-theoretic renormalization group 
techniques. 
Monte Carlo simulations are employed to determine the new exponents 
in dimensions $d \leq 3$.
\end{abstract}

]


Nonequilibrium critical phenomena have been the focus of intense
theoretical studies in recent years, prominent examples being phase
transitions in driven diffusive systems \cite{dridif}, kinetic Ising
models \cite{kinism}, nonequilibrium growth models \cite{gromod},
diffusion-limited chemical reactions \cite{chemrc,jcardy}, 
percolation-like processes \cite{percol}, and generally in lattice
models or cellular automata.  
The concept of only a few distinct universality classes which
determine the critical exponents has been remarkably successful for
the description of static and even dynamic critical phenomena near 
equilibrium phase transitions \cite{crtdyn}. 
However, it is still an unresolved issue whether a similarly small
number of universality classes can be identified in nonequilibrium 
situations, which potentially contain much richer behavior.
In addition to the dynamic universality classes listed in
Ref.~\cite{crtdyn}, some notable new candidates have emerged in quite
different circumstances, for example, the roughening transition in the
KPZ equation \cite{kapezt}, and the critical point of directed
percolation (DP) \cite{dirper}, as described by Reggeon field theory
\cite{regfth}.
In fact, the DP universality class appears to cover the majority of
phase transitions from non-trivial active into absorbing states where
the order parameter noise vanishes.
However one striking exception to this rule occurs when the relevant
dynamic processes are constrained by an additional local ``parity''
symmetry, in which case the universality class is that of branching
and annihilating random walks with an even number of offspring
\cite{evbarw,cartau}.

This work is originally motivated by recent studies of a nonequilibrium 
growth model for adsorption and desorption of particles which displays 
a roughening transition in $d = 1$ \cite{hinric}. 
The key feature is that desorption may take place only at the edge of
an existing plateau so that particles cannot be desorbed from a
completed layer. 
Because of this property the dynamic processes at a given height level
decouple from all higher levels, i.e. the processes at different
heights are unidirectionally coupled. 
This growth model can in fact be related to a reaction-diffusion
problem for a hierarchy of particle species $A, B, \ldots$, which 
correspond to different height levels, respectively.
Each of these species is subject to the prototypical DP processes 
(see Refs.~\cite{regfth,cartau})
\begin{equation}
        A \to A + A \ , \quad A + A \to A \ , \quad A \to \emptyset \ ,  
 \label{dpproc}
\end{equation}
with reaction rates $\sigma_A$, $\lambda_A$, and $\mu_A$, respectively
(and accordingly for particles $B, \ldots$). 
However, there is also a coupling from each previous hierarchy level
to the next one via random particle transmutations
\begin{equation}
        A \to B \ , \quad B \to C \ , \quad \ldots
 \label{coupdp}
\end{equation}  
with rates $\mu_{AB}, \mu_{BC}, \ldots$, but {\em no} feedback
mechanism from the $B$ to the $A$ species etc. 
For simplicity, we shall assume that the diffusion constant $D$ for
{\em all} particle species $A, B, \ldots$ is identical.
When $\mu_{AB} = \mu_{BC} = \ldots = 0$, the system decouples, and for
the $i$-th species there is a critical point separating an active from
an absorbing state, where essentially the branching rate $\sigma_i$
and decay rate $\mu_i$ balance one another. 
At this point there are {\em three} independent critical exponents 
governing ({\it i}) the divergence of the correlation length 
$\xi \propto |r_i|^{-\nu_\perp}$ (where $r_i \propto \mu_i - \sigma_i$
denotes the deviation from the critical point), ({\it ii}) the
critical slowing down of characteristic time scales 
$t_c \propto \xi^z \propto |r_i|^{-\nu_\parallel}$, with 
$\nu_\parallel = z \nu_\perp$, and ({\it iii}) the value of the 
asymptotic particle density in the active phase, 
$n_i(t \to \infty) \propto |r_i|^\beta$. 
Alternatively, this third independent exponent may be swapped for
$\eta_\perp$, which characterizes how the equal-time pair correlation
function decays at the critical point $r_i = 0$, 
$G(|{\bf x}|) \propto 1/|{\bf x}|^{d + z - 2 + \eta_\perp}$ 
[see Eq.~(\ref{dpbeta}) below].  
These exponents should be those of the DP universality class, with
upper critical (spatial) dimension $d_c = 4$.
Numerical simulations in $d = 1$ have confirmed this, and furthermore
revealed that $\nu_\perp \approx 1.1$ and $z \approx 1.6$ remain
unchanged at {\em all} hierarchy levels even when the rates 
$\mu_{AB} \ldots$ are switched on.  
However, while $\beta_A = \beta^{(1)} \approx 0.27$ as in DP for the
primary $A$ particles, the particle density exponents for
secondary and higher hierarchy levels are considerably reduced to 
$\beta_B = \beta^{(2)} \approx 0.11$, and $\beta^{(3)} \approx 0.04$,
{\em provided} the critical points for the processes on the different
levels coincide \cite{hinric}.

In the following, we shall mainly study the two-level hierarchy of DP 
processes (\ref{dpproc}) for $A$ and $B$ particles coupled by the
reaction $A \to B$.
We explain the reduced value of $\beta^{(2)}$ as a consequence of the 
{\em multicritical behavior} that emerges when $r_A = r_B = r \to 0$.
It appears then rather likely that similar multicritical behavior
should emerge in more general contexts than the specific growth model
and diffusion-limited chemical reactions (\ref{dpproc}),
(\ref{coupdp}); namely whenever DP-like processes are coupled
unidirectionally {\em without} feedback. 
A qualitatively correct description is obtained by analyzing the 
corresponding mean-field rate equations, which yield
$\beta^{(k)} = 1/2^{k-1}$ at the multicritical point on the $k$-th
hierarchy level.
With the aid of the renormalization group (RG), we then identify an 
additional {\em independent} crossover exponent $\phi$ related to the 
relevant scaling field ($\mu_{AB} / D$) of the two-level hierarchy
multicritical point.
A scaling relation will be derived expressing $\beta^{(2)}$ in terms 
of $\phi$ and the conventional DP exponent $\beta = \beta^{(1)}$.
Using a field-theoretic representation (see Refs.~\cite{masfth,redfth}) 
based on the master equation for this 
stochastic process, we furthermore compute the fluctuation 
corrections to $\phi$ and $\beta^{(2)}$ to first order in 
$\epsilon = 4 - d$ \cite{yadin}. Finally, these exponents are determined 
accurately in $d \leq 3$ dimensions by means of Monte Carlo simulations.

\begin{table}
\begin{tabular}{|c||l|l|l|l|}
               & $d=1$        & $d=2$     & $d=3$    & $d=4-\epsilon$
\\\hline
$\beta^{(1)}$  & $0.271(10)$  & $0.57(5)$ & $0.78(7)$  
& $1-\epsilon/6+ O(\epsilon^2) $\\
$\beta^{(2)}$  & $0.108(10)$  & $0.30(4)$ & $0.35(6)$     
& $1/2 - \epsilon / 6 + O(\epsilon^2) $ \\
$\beta^{(3)}$  & $0.038(8)$   & $0.13(3)$ & $0.15(4)$     
& $1/4 - \epsilon / 6 + O(\epsilon^2) $ \\
\end{tabular}
\caption{Results for the particle density exponents $\beta^{(k)}$
obtained from Monte Carlo simulations and RG calculations.}
\end{table}

We start by writing down the mean-field rate equations for the average
local densities $n_A({\bf x},t)$ and $n_B({\bf x},t)$ of the $A$ and
$B$ particles, both subject to the reactions (\ref{dpproc}), and
coupled via the random particle transmutation $A \to B$,
\begin{eqnarray}
        \partial_t n_A &&= D (\nabla^2 - r_A) n_A - \lambda_A n_A^2 \ ,  
 \label{mfreqa} \\ 
        \partial_t n_B &&= D (\nabla^2 - r_B) n_B  - \lambda_B n_B^2 
                                                   + \mu_{AB} n_A  \ ,
 \label{mfreqb}
\end{eqnarray}
where $r_A = (\mu_A + \mu_{AB} - \sigma_A) / D$ and 
$r_B = (\mu_B - \sigma_B) / D$.
Considering first the equation for $n_A$, it is clear that for 
$r_A > 0$ the only stationary state is $n_A = 0$ (inactive phase). 
On the other hand, if $r_A < 0$ the density will asymptotically
approach $n_A^\infty = D (-r_A) / \lambda_A > 0$ (active state), and
hence $\beta^{(1)} = 1$ in this approximation. From 
the diffusive character of Eq.~(\ref{mfreqa}), and also from the 
functional form of its linear term, it follows that $z = 2$ and 
$\nu_\perp = 1/2$ in mean-field. 
These last two results remain intact for the $B$ particles as well, even
in the presence of particle influx via the reaction $A \to B$. 
However, the possible stationary states for $n_B$ actually depend
on the value of $r_A$ as well as on $r_B$, if $\mu_{AB} > 0$.
If $r_A > 0$, and therefore $n_A(t \to \infty) = 0$,
Eq.~(\ref{mfreqb}) reduces to a mean-field DP process with critical
point $r_B = 0$ and $\beta = 1$. 
On the other hand, in the $A$-particle active state, $r_A < 0$, the  
stationary solution of (\ref{mfreqb}) becomes $n_B^\infty = 
[(D r_B / 2 \lambda_B)^2 + \mu_{AB} n_A^\infty / \lambda_B]^{1/2} 
- D r_B / 2 \lambda_B$; i.e., there is merely a crossover from a
low-density to a high-density regime in the vicinity of $r_B = 0$.
An interesting situation arises when $r_A = r_B = r$, as in this case
the second term in the square brackets dominates for $|r| \to 0$,
\begin{equation}
        n_B^\infty = [D \mu_{AB} (-r)/ \lambda_A \lambda_B]^{1/2} 
                        + O(|r|) \ .
 \label{mfdens}
\end{equation}
Hence $\beta^{(2)} = 1/2$ instead of the mean-field DP value. 
More generally, this multicritical behavior arises in the regime 
$(D r_B / 2 \lambda_B)^2 \ll D (-r_A) \mu_{AB} / \lambda_A \lambda_B$
for $r_A \uparrow 0$. 
On the other hand, for $r_B > 0$ and 
$(D r_B / 2 \lambda_B)^2 \gg D (-r_A) \mu_{AB} / \lambda_A \lambda_B$
one finds $n_B^\infty = (-r_A) \mu_{AB} / r_B \lambda_A$, and thus 
one expects DP transitions for the $B$ species {\em both} as $r_B \to 0$
(with $r_A > 0$) {\em and} $r_A \to 0$ ($r_B > 0$) \cite{future}. 

Notice that there are three critical lines, namely ($r_A = 0$, 
$r_B > 0$), ($r_A = 0$, $r_B < 0$), and ($r_A > 0$, $r_B = 0$), two of
which are critical lines for the $B$ particles, meeting at the point 
$r_A = r_B = 0$. 
This special point can therefore be interpreted as a multicritical
point, with the mean-field order parameter exponent halved as compared
to the ordinary critical point \cite{eqtric}.
Clearly the generalization of Eqs.~(\ref{mfreqa}), (\ref{mfreqb}) to 
$k$ hierarchy levels leads to $\beta^{(k)} = 1/2^{k-1}$ at the 
multicritical point $r_A = r_B = \ldots = 0$.

For $d < d_c = 4$, one expects that these mean-field values will be
strongly modified by fluctuation effects.
We have performed Monte Carlo simulations in $d \leq 3$ dimensions in
order to determine the survival probability exponent 
$\beta^{(k)}/\nu_\parallel^{(k)}$ up to the third hierarchy level. 
The DP processes (\ref{dpproc}) are realized by a cellular automaton
which evolves by parallel updates in which a site at time $t+1$
becomes active with probability $p$ provided that the same site, or
at least one of its nearest neighbors, was active at time $t$.
The coupling (\ref{coupdp}) between levels is incorporated by the rule
that active sites at a given level simultaneously impose active sites on
all higher levels. 
We perform dynamic Monte Carlo simulations \cite{grassb} where one
measures the temporal evolution of the system  at criticality
starting from a single seed (a single $A$ particle).
Assuming that $\nu_\parallel^{(k)} = \nu_\parallel$ is identical on all 
levels $k$, we can use the survival probability to determine the
density exponents, giving in $d = 1$ $\beta^{(1)} = 0.271(10)$ as 
in DP \cite{jensen}, $\beta^{(2)} = 0.108(10)$, and $\beta^{(3)} = 0.038(8)$. 
These results are consistent with direct estimations of $\beta^{(k)}$ 
obtained in off-critical simulations \cite{hinric}.
The mean-square spreading exponents yield the exponent $z$, with the
results $z^{(1)} = 1.57(2)$ (DP), $z^{(2)} = 1.58(2)$ and 
$z^{(3)} = 1.59(2)$, which confirms that these exponents are
identical on all levels. 
Similar results are obtained in $d = 2$, $d = 3$ (see Table I).

In order to better understand the very low values found for
$\beta^{(k)}$ in the simulations, we clearly have to take
fluctuation effects into account. In low dimensions, where
reaction noise in the processes (\ref{dpproc}), (\ref{coupdp}) 
is highly important, a straightforward Langevin-type description
of the noise can be dangerous, because of the highly non-trivial
form of the noise correlations. 
Instead, it is useful to start from the corresponding master equation,
and utilize a standard formalism involving a second-quantized bosonic
operator representation to derive an effective field theory directly
\cite{masfth,redfth}. 
It is important to note that apart from the continuum limit, this
procedure is {\em exact} and requires no additional assumptions
regarding the precise form of the noise.
In the present case, the field theory describing the two-level coupled
reactions in $d$ dimensions is defined by the action (we omit terms
related to the initial state)
\begin{eqnarray}
        S &&= \int \! d^dx \int \! dt \Bigl\{
        {\bar a} \left[ \partial_t + D (r_A - \nabla^2) \right] a
        - \sigma_A {\bar a}^2 a \nonumber \\
        &&+ \lambda_A \left( {\bar a} a^2 + {\bar a}^2 a^2 \right)
        - \mu_{AB} {\bar b} a \\
 \label{fthact} 
        &&+ {\bar b} \left[ \partial_t + D (r_B - \nabla^2) \right] b
        - \sigma_B {\bar b}^2 b 
        + \lambda_B \left( {\bar b} b^2 + {\bar b}^2 b^2 \right) 
        \Bigr\} \ . \nonumber
\end{eqnarray}
We remark that $\langle a \rangle = n_A$ and 
$\langle b \rangle = n_B$, whereas the fields  
${\bar a}, {\bar b}$ have no direct physical interpretation. 

It is now convenient to rescale the fields according to 
${\bar a} = (\lambda_A / \sigma_A)^{1/2} {\bar \psi}_0$,
$a = (\sigma_A / \lambda_A)^{1/2} \psi_0$,
${\bar b} = (\lambda_B / \sigma_B)^{1/2} {\bar \varphi}_0$,
$b = (\sigma_B / \lambda_B)^{1/2} \varphi_0$, and also define
new couplings $u_0 = 2 (\sigma_A \lambda_A)^{1/2}$, 
$u_0' = 2 (\sigma_B \lambda_B)^{1/2}$, as well as
$\mu_0 = \mu_{AB} (\sigma_A \lambda_B / \sigma_B \lambda_A)^{1/2}$
(henceforth, the subscript ``0'' denotes unrenormalized
quantities).
If we introduce a length scale $\kappa^{-1}$ and correspondingly
measure times in units of $\kappa^{-2}$ (i.e., $[D_0] = \kappa^0$),
we find that the new fields have scaling dimension $\kappa^{d/2}$,
while $[r_A] = [r_B] = [\mu_0] = \kappa^2$, which are thus relevant
perturbations in the RG sense.
On the other hand, $[u_0] = [u_0'] = \kappa^{2-d/2}$, and the
corresponding DP nonlinearities become marginal in $d_c = 4$
dimensions, as expected \cite{regfth}. 
It is important to note, however, that 
$[\lambda_A] = [\lambda_B] = \kappa^{2-d}$, and hence these couplings 
are {\em irrelevant} as compared to $u_0$ and $u_0'$, and 
may be omitted in the effective action (see Ref.~\cite{cartau}), 
which consequently reads 
\begin{eqnarray}
        S_{\rm eff} &&= \int \! d^dx \int \! dt \Bigl\{ 
        {\bar \psi}_0 \left[ \partial_t + D_0 (r_A - \nabla^2) 
                                \right] \psi_0 \nonumber \\
        &&- {u_0 \over 2} \left( {\bar \psi}_0^2 \psi_0 
                        - {\bar \psi}_0 \psi_0^2 \right) 
                - \mu_0 {\bar \varphi}_0 \psi_0 \label{effact} \\
        &&+ {\bar \varphi}_0 \left[ \partial_t + D_0 (r_B - \nabla^2) 
                                \right] \varphi_0 
        - {u_0' \over 2} \left( {\bar \varphi}_0^2 \varphi_0 
                - {\bar \varphi}_0 \varphi_0^2 \right) \Bigr\} \ . 
        \nonumber
\end{eqnarray}
The saddle-point equations corresponding to this action are precisely
the mean-field rate equations (\ref{mfreqa}), (\ref{mfreqb}) but with
$\psi_0$, $\varphi_0$ instead of $n_A, n_B$, and with $\mu_{AB}$,
$\lambda_A$, and $\lambda_B$ replaced by $\mu$, $u_0/2$, and $u_0'/2$.
We remark that the harmonic part of this action can be diagonalized 
{\em only} when $r_A \not= r_B$, leading to new cubic vertices mixing
the fields $\psi_0$ and $\varphi_0$ [compare Eq.~(\ref{addver}) below]. 
However, these additional vertices have no influence on the behavior
of the $A$ and $B$ species \cite{jansen}, and the diagonalized theory
leads to the anticipated DP critical behavior at both critical lines
($r_A = 0$, $r_B \not= 0$) and ($r_A > 0$, $r_B = 0$) \cite{future}. 

Here we are predominantly interested in the multicritical point 
$r_A = r_B = r_0 \to 0$, and thus we must work with the non-diagonal
action (\ref{effact}). 
Furthermore, due to the inclusion of the relevant scaling field
$\mu_0$, a number of ``mixed'' cubic vertices are additionally
generated at the tree level, all of which have scaling dimension
$\kappa^{2-d/2}$.
These vertices must be taken into account separately, and hence we
need to replace (\ref{effact}) with 
$S_{\rm mc} = S_{\rm eff} + \Delta S$, where 
\begin{eqnarray}
        \Delta S =  \int \! d^dx \int \! dt 
        \Bigl[ &&- s_0 {\bar \varphi}_0 {\bar \psi}_0 \psi_0
        - {s_0' \over 2} {\bar \varphi}_0^2 \psi_0 \nonumber \\
        &&+ {{\tilde s_0} \over 2} {\bar \varphi}_0 \psi_0^2
        + {\tilde s}_0' {\bar \varphi}_0 \varphi_0 \psi_0 \Bigr] \ .
 \label{addver}
\end{eqnarray}
It is now a straightforward task to compute the renormalized
couplings and the RG fixed points to one-loop order. 
The DP nonlinearities in (\ref{effact}) remain unaffected by the
``mixed'' vertices (\ref{addver}), and the stable nontrivial fixed
points for the associated dimensionless renormalized couplings 
$u = u_0 A_d^{1/2} \kappa^{(d-4)/2}$ and 
$u' = u_0' A_d^{1/2} \kappa^{(d-4)/2}$, where 
$A_d = \Gamma(3 - d/2) / 2^{d-1} \pi^{d/2}$, are 
\begin{equation}
        [(u/D)^*]^2 = [(u'/D)^*]^2 = 4 \epsilon/3 + O(\epsilon^2) \ .
 \label{dpfxpt}
\end{equation}
Hence the critical exponents 
$\eta_\perp = - \epsilon / 12 + O(\epsilon^2)$, 
$\nu_\perp^{-1} = 2 - \epsilon / 4 + O(\epsilon^2)$, and 
$z = 2 - \epsilon / 12 + O(\epsilon^2)$ remain those of the DP
universality class at {\it all} hierarchy levels, whereas the exponent
$\beta^{(k)}$ is unchanged only at the first level
\begin{equation}
        \beta^{(1)} = {\nu_\perp \over 2} (d + z - 2 + \eta_\perp) =
        1 - \epsilon / 6 + O(\epsilon^2) \ .
 \label{dpbeta}
\end{equation}
For the similarly defined renormalized ``mixed'' cubic vertices, one
finds two RG {\em fixed lines}. 
The first is given by
\begin{eqnarray}
        &&(s/D)^* = - ({\tilde s}'/D)^* , \,
        ({\tilde s}/D)^* - (s'/D)^* = 2 (s/D)^* , 
 \label{fline1} \\
        &&\qquad 
        [(s/D)^*]^2 = 2 (\epsilon/3)^{1/2} [(s/D)^* + (s'/D)^*] \ ,
 \label{connec}
\end{eqnarray}
with stability matrix eigenvalues $0, 0, -\epsilon/3, -\epsilon/3$,
which imply that this fixed line (including the Gaussian fixed point
for $\Delta S$) is unstable for $d < 4$.
The second fixed line,
\begin{equation}
        (s'/D)^* = ({\tilde s}/D)^* , \,
        (s/D)^* + ({\tilde s}'/D)^* = 2 (\epsilon/3)^{1/2} ,
 \label{fline2}
\end{equation}
with again Eq.~(\ref{connec}), turns out to be {\em stable} as its
stability matrix has eigenvalues 
$0, \epsilon/3, \epsilon/3, 4 \epsilon/3$ \cite{future}.  
We remark that both fixed lines (\ref{fline1}) and (\ref{fline2}) with
(\ref{connec}) satisfy the condition $(\epsilon/3)^{1/2} 
[(s'/D)^* + ({\tilde s}/D)^*] = - (s/D)^* ({\tilde s}'/D)^*$, 
which ensures the cancellation of strongly singular (UV-divergent in
$d = 2$) diagrams for the renormalization of $\mu_0$.
The RG eigenvalue of $\mu$ at these fixed lines then becomes
$y_\mu = 2 + \epsilon / 6$, and $y_\mu = 2 - \epsilon / 6$,
respectively.
In principle a product of quartic vertices and $\mu_0$ might enter the
renormalizations of the three-point functions as well. 
However, we have checked that these additional couplings all have
negative RG eigenvalues and are therefore irrelevant \cite{future}.

We next define the crossover exponent $\phi$ related to the new
scaling field $\mu / D$, as it appears in the general scaling form for
the $B$--species ``density'' field, 
\begin{equation}
        \varphi(|r|,\mu/D) = |r|^{\beta^{(1)}}
                \, {\hat \varphi}(|r|^{-\phi} \, \mu / D)
 \label{densca}
\end{equation}
and identify
\begin{equation}
        \phi = \nu_\perp (y_\mu + z - 2) = 1 + O(\epsilon^2)
 \label{1lpchi}
\end{equation}
at the stable fixed line.

Finally, we relate the above exponent $\phi$ to the density
exponent $\beta^{(2)}$ on the second hierarchy level. 
We first use the RG flow equations for the scale-dependent
renormalized parameters, and then match to our earlier mean-field
result (\ref{mfdens}) for the density in the active phase in the
vicinity of the multicritical point. 
In terms of the rescaled fields, we found
$\varphi_0^\infty = (2\mu_0 \psi_0^\infty / u_0')^{1/2}$ for 
$r_0 \to 0$.
Essentially, we now have to replace the bare parameters here by their 
flowing renormalized counterparts, when inverse length scales are
rescaled according to $\kappa \to \kappa \ell$.
To that end, we first note that the fields $\psi$ and $\varphi$
scale as $\ell^{(d + z - 2 + \eta_\perp)/2}$ [which, with the matching
condition $\ell = (-r)^{\nu_\perp}$, actually implies the scaling
relation cited in Eq.~(\ref{dpbeta})]. 
Furthermore, the $O(\epsilon)$ fixed points (\ref{dpfxpt}) require 
$u^{(')}(\ell) / D(\ell) \to {\rm const.}$, and according to our
definitions, we have $r(\ell) \to r \ell^{-1/\nu_\perp}$ and
$\mu(\ell) / D(\ell) \propto \ell^{- \phi / \nu_\perp}$. 
Upon identifying $\ell = (-r)^{\nu_\perp}$, we finally arrive at
\begin{equation}
        \beta^{(2)} = \beta^{(1)} - \phi / 2 
                = 1/2 - \epsilon / 6 + O(\epsilon^2) \ ,
 \label{tpbeta}
\end{equation}
using our above results \cite{yadin}. 
Equivalently, this follows by demanding that the scaling function
in Eq.~(\ref{densca}) behave as ${\hat \varphi}(x) \to x^{1/2}$ for
$x \to \infty$, as prescribed by mean-field theory (\ref{mfdens}).
The $O(\epsilon)$ result (\ref{tpbeta}) constitutes a sizable 
{\em downward} renormalization through fluctuation effects, as is also
evident from the simulation data in Table I.
These can be explained with the values $\phi \approx 0.33$, $0.54$,
and $0.86$ in $d = 1$, $2$, and $3$, respectively.

In principle, these considerations are readily generalized to further
hierarchy levels \cite{future}. 
For example, although a combination of processes $A \to B$ and 
$B \to C$ will generate the reaction $A \to C$ as well, one finds that
the additional relevant scaling field $\mu_{AC}$ does not enter the
leading contribution to the $C$ density near the multicritical point 
$r_A = r_B = r_C = r \to 0$. Hence $n_C \approx 
[D \mu_{AB} \mu_{BC}^2 (-r) / \lambda_A \lambda_B \lambda_C^2]^{1/4}$,
implying that no additional independent exponents need to be
introduced. We then find $\beta^{(3)} = \beta^{(1)} - 3 \phi / 4 
= 1/4 - \epsilon / 6 + O(\epsilon^2)$.

In summary, we have studied a hierarchy of {\em unidirectionally
coupled} DP processes (no feedback), as defined by the
diffusion-limited reactions (\ref{dpproc}), (\ref{coupdp}). 
Already within mean-field theory, we have identified a purely 
{\em dynamic multicritical point}, which occurs when the DP critical 
points at each level coincide. 
Using a field-theoretic representation, we have evaluated to 
$O(\epsilon)$ the crossover exponent $\phi$ related to the relevant 
scaling field $\mu_{AB} / D$. 
We then derived a scaling relation which allowed us to determine
$\beta^{(2)}$. 
In accord with our Monte Carlo simulation results, fluctuation effects
lead to a strong {\em downward} renormalization of the density
exponents as compared with the mean-field values 
$\beta^{(k)} = 1/2^{k-1}$. We emphasize that
the existence of such a multicritical regime is to be expected 
generically whenever processes of the ubiquitous DP universality class 
are coupled unidirectionally.

We benefited from discussions with J.L.~Cardy, M.R.~Evans,
Y.Y.~Goldschmidt, and D.~Mukamel. 
U.C.T. acknowledges support from the European Commission, contract 
No.~ERB FMBI-CT96-1189, and from the Deutsche Forschungsgemeinschaft, 
Gz.~Ta 177 / 2.

$^*$ Present address: Institut f\"ur Theoretische Physik, 
Physik-Department der TU M\"unchen, James-Franck-Stra\ss e, 
D-85747 Garching, Germany.



\begin{thebibliography}{99}
\bibitem{eqtric} This is remarkably similar to the equilibrium
                tricritical behavior in the 
                ($r \Phi^2 + u \Phi^4 + v \Phi^6$)-model,
                where $\beta = 1/2$ for $u > 0$ and $\beta = 1/4$ for
                $u = 0$ in the Landau approximation. 
                Notice, however, that the mechanism leading to the
                reduction of the exponent $\beta$ is quite different
                here.
\end{thebibliography}
\end{document}